# UV photo-dissociation of methyl-bromide and methyl-bromide cation studied by velocity map imaging

Valérie Blanchet[2,3], Peter C. Samartzis[1], Alec M. Wodtke[1]

[1]Department of Chemistry and Biochemistry,
University of California, Santa Barbara, CA 93106, USA

[2]Université de Toulouse ; UPS ; Laboratoire Collisions Agrégats Réactivité, IRSAMC ; F-31062 Toulouse, France
[3]CNRS ; UMR 5589 ; F-31062 Toulouse, France

**Abstract**

We employ the velocity map imaging technique to measure kinetic energy and angular distributions of state selected $CH_3$ ($v_2$=0,1,2,3) and Br ($^2P_{3/2}$,$^2P_{1/2}$) photofragments produced by methyl bromide photolysis at 215.9 nm. These results show unambiguously that the Br and Br* forming channels result in different vibrational excitation of the umbrella mode of the methyl fragment. Low energy structured features appear on the images which arise from $CH_3Br^+$ photodissociation near 330 nm. The excess energy of the probe laser photon is channeled into $CH_3^+$ vibrational excitation, most probably in the $\nu_4$ degenerate bend.

PACS: 82.50.Hp, 33.80.Gj, 33.15.Fm

Keywords: methyl bromide, photodissociation

## I INTRODUCTION

As in all methyl halides or alkyl and aryl halides, the $\tilde{A} \leftarrow X$ band of $CH_3Br$, characterized by excitation to the anti-bonding σ* orbital localized along the C-Br bond, exhibits a broad photodissociation continuum. This diffuse band in the ultraviolet range, environmentally relevant to atmospheric chemistry,[1,2] is characterized by a prompt dissociation via two dominant photofragmentation pathways:

$$\begin{aligned} \mathrm{CH_3Br} + \hbar\nu \rightarrow\ & \mathrm{CH_3} + \mathrm{Br}\left({}^2\mathrm{P}_{3/2}\right) \\ & \mathrm{CH_3} + \mathrm{Br}^*\left({}^2\mathrm{P}_{1/2}\right) \end{aligned}$$

The spin-orbit coupling of the halide lifts degeneracy to give rise to three optically allowed electronic transitions to states which are, in order of increasing energy, the ${}^3Q_{1,}$ ${}^3Q_0$ and ${}^1Q_1$ states. In the $C_{3v}$ geometry, dissociation via the ${}^XQ_1$ (with X=1 or 3) and ${}^3Q_0$ states correlates adiabatically to $Br\left({}^2P_{3/2}\right)$ and $Br^*\left({}^2P_{1/2}\right)$, respectively. These transitions can be identified through the orientation of the transition electric dipole moment relative to the C-Br bond: the ${}^XQ_1$ transitions are perpendicular while the ${}^3Q_0$ transition is parallel. The dipole moments of ${}^XQ_1 \leftarrow \tilde{X}$ arise from intensity borrowing from Rydberg states lying ~1 eV higher in energy,[3] similarly to $CH_3I$.[4] Near the absorption maximum of the $\tilde{A} \leftarrow X$ band, perpendicular and parallel transitions of similar probability are allowed,[5] in contrast to the $\tilde{A} \leftarrow X$ band in $CH_3I$. The Franck-Condon region is centered around an equilibrium C-Br bond distance of 1.95±0.1 Å.[6,7] Breaking $C_{3v}$ symmetry, induced for instance by the rocking mode $\nu_6$, leads to coupling between ${}^1Q_1$ and ${}^3Q_0$ surfaces opening possible non-adiabatic transitions that strongly influence the product branching ratio between the two spin-orbit channels. This can be quite different for alkyl halides of $C_s$ symmetry for which in some cases no avoided crossing emerges, for example in $CH_2ClBr$[8] or in $C_3H_7Br$.[9,10] The curve crossing in $CH_3Br$ has been calculated to occur at a C-Br internuclear distance around 2.445 Å.[6] Its probability has been experimentally determined.[5] In all these systems, the $^3Q_1$ state seems to be decoupled from the $^3Q_0$ state.[11] Many aspects of this photodissociation system have been summarized before.[12]

A study carried out around the absorption maximum of the $\tilde{A} \leftarrow X$ band at 202 nm and analyzing the kinetic energy distribution of Br fragments suggested that the internal energy of $CH_3$ is imparted into the $\nu_2$ umbrella mode and that the vibrational excitation was different for the two spin-orbit channels.[13] The $Br\left({}^2P_{3/2}\right)$-producing channel appears to peak around $v_2$=3, while the $Br^*\left({}^2P_{1/2}\right)$- producing channel peaks around $v_2$=1 or 2.[13] Similar conclusions were reached through use of photofragment imaging of product Br-atoms ($\lambda_{photolysis}$ ~ 205 nm)[14] and of $CH_3$ by non-resonant multiphoton ionization ($\lambda_{photolysis}$ ~ 226 to 218 nm).[5] Despite a lack of state specific detection of $CH_3$, these studies suggest that the umbrella mode vibrational distribution of $CH_3$ follows the general rules that are $\upsilon_2^{\max} > 0$ and $\upsilon_2^{\max}\left(Br\right) > \upsilon_2^{\max}\left(Br^*\right)$.

Similar conclusions have been derived from observations in other molecules, such as $CH_3I$, $CF_3Br$ and $CH_3Cl$.[12,15] As a function of the photolysis energy, the vibrational distribution of the umbrella mode is not expected to change drastically.[16] Methyl radical has a planar equilibrium geometry, while $CH_3$ in the ground electronic state of $CH_3Br$ has a pyramidal geometry with an equilibrium angle around 111.5°.[6,7] In the case of dissociation of $CH_3I$ in the $\tilde{A} \leftarrow \tilde{X}$ band, the vibrational distribution is frequently rationalized by an abrupt change from pyramidal to planar geometry at the seam of the $^1Q_1$ - $^3Q_0$ curve crossing, so that any trajectory going through the non-adiabatic transition, which leads to Br ($^2P_{3/2}$), receives extra umbrella mode excitation.[11] In other words, if the Br producing channel is populated significantly by this non-adiabatic transition, its umbrella activity is expected to be more excited than that of the Br* producing channel for which the pyramidal to planar relaxation occurs more gradually and adiabatically along the $^3Q_0$ surface.

Another suggested explanation not involving the $^1Q_1$-$^3Q_0$ curve crossing relies on the difference in slope of the potential energy surface of $^1Q_1$ and $^3Q_0$ states as a function of the distance between C and the center of mass of the three H atoms in the vicinity of the Franck-Condon region. The $^1Q_1$ surface is steeper, therefore the umbrella excitation will have less time to relax from pyramidal to planar geometry compared to trajectories following the $^3Q_0$ state.[14,16,17] In all these studies, with the exception of some methyl substituted bromides,[18] it has been pointed out that the methyl fragment is likely to be rotationally cold due to a weakly bent excited state for both surfaces. This is confirmed by studies of the photodissociation of rotationally state-selected methyl bromide between 213 nm and 235 nm.[19] The rotational distribution of $CH_3$ photofragments extends up to N=6 for photolysis at 230 nm and slightly higher - N=9 - when photolysis is performed at 212.8nm, with almost no difference between a Br* and Br channels.[19] This experiment has investigated mainly the production of vibrational ground state methyl and the influence of deuteration. Similar work has been performed on the extreme red wing (240-280 nm) of the $\tilde{A} \leftarrow \tilde{X}$ absorption band with an effusive molecular beam.[20]

We report here kinetic energy and angular distributions of bromine and methyl fragments recorded by the velocity map imaging technique following photodissociation of $CH_3Br$ at 215.9 nm. The vibrational states of methyl radicals were probed by resonant enhanced multiphoton ionization (2+1 REMPI) via the intermediate Rydberg $3p\ ^2A_2''$ state. At this photolysis energy the direct absorption ratio of parallel to perpendicular excitation is around $\left[{}^3Q_0\right]/\left[{}^3Q_1 + {}^1Q_1\right]$=1.5.[5] We have recorded velocity map images of individual vibrational states of $CH_3$ excited in the $\nu_2$ umbrella mode up to $\nu_2$=3, in contrast to previous experiments in which only the kinetic energy distribution of Br fragments were available[14,20] or where $CH_3$ was detected only in its ground vibrational state[19] or without vibrational state selection.[5] The images show two features that can be assigned to formation of Br and $Br^*$. We are thus able to unambiguously confirm the prior interpretation of Br atom velocity maps and photofragment images which implicated a greater degree of $\nu_2$-vibrational excitation in the Br-forming channel.

We also observe for the first time, the photolysis of $CH_3Br^+$ cation in the near UV. This results in a highly structured velocity map image, which can be assigned to progression of high vibrational states of the $CH_3^+$ product ion. An indirect photodissociation mechanism involving non-adiabatic coupling is proposed.

## II EXPERIMENTAL SET-UP

Methyl bromide with 99% purity containing the natural abundance of bromine isotopes was used without further purification. A 10% $CH_3Br$/He mixture at 0.5 bar stagnation pressure expanded supersonically through a solenoid pulsed valve (General Valve Series 9, Parker Hannifin Corp.) with a 0.8 mm orifice to produce a 300 μs long pulsed molecular beam. After passing through a skimmer, the molecular beam was collimated to 1 mm and entered the ion optical region of a standard velocity mapping ion source.[21] The molecular beam was then intersected midway between the repeller and the extractor plates by two counter propagating laser beams both perpendicular to the time-of flight axis, which dissociated $CH_3Br$ and selectively ionized the fragment of choice. The ions produced were accelerated before being mass-selected at the end of a 1-m long time-of-flight tube by gating the gain of a 7.5 cm diameter dual microchannel plate (MCP)/phosphor imaging detector (Burle ElectroOptics). The gating voltage pulse was applied on the front MCP-plate and was typically -500 V over 170 ns. The images were recorded typically over 36000 laser shots by imaging software (Davis, LaVision) on a (640x480) CCD camera (XC7500, Sony) coupled to a frame grabber (PCimage SG, Matrix Vision). The repeller voltage was fixed at 5kV for the methyl fragments so that the detector surface is 80% filled and at 3kV for the Br fragment to enlarge the images. The optimal velocity mapping condition corresponds to an extractor/repeller voltage ratio of 0.7. In the kinetic energy range explored here, one pixel on the image corresponds roughly to 30 meV shift.

The repetition rate is fixed at 10 Hz by the photodissociation laser. The 215.9 nm photolysis beam was produced by frequency doubling the output of an Nd:YAG (Continuum Powerlite 7010, 3$^{rd}$ harmonic at 355 nm) pumped dye laser (Sirah) operating on Stilbene, through a 7 mm BBO crystal. The 1 mJ/pulse output was focused by a 50 cm focal length lens on the molecular beam. The photofragments were selectively ionized by a REMPI probe laser. $Br\left({}^2P_{3/2}\right)$ photofragment was the only species probed by a single laser scheme; namely, photodissociation and detection by (2+1) REMPI via the ${}^4D^0_{5/2}$ state at 215.9 nm. The methyl photofragments were photoionized by (2+1) REMPI via the $3p\ {}^2\mathrm{A}_2^{"}$ Rydberg state,[22,23] using the doubled (circa 340 nm) output of a dye laser (Continuum ND6000) operating on a mixture of DCM/LDS698 and pumped by the second harmonic of an Nd:YAG laser (Continuum Powerlite 8020). About 4 mJ/pulse of this probe laser light was focused by a 60 cm focal length lens onto the molecular beam. Vibrationally excited states of $CH_3$ were probed via $2^v_v$ two-photon transition, up to v=3. The REMPI laser for each of those transitions was tuned to the transition maximum, which corresponds to a Q branch. The $Br^*\left({}^2P_{1/2}\right)$ photofragments were photoionized by (2+1) REMPI via the ${}^2P^0_{1/2}$ state at 238.6 nm. To produce this wavelength, the doubled output of a dye laser (Continuum ND6000) operating with R640 and pumped by the second harmonic of an Nd:YAG laser (Continuum Powerlite 8020) was mixed with the pump laser’s fundamental of 1064 nm.

In the case of methyl ion detection, no ion background was observed with the REMPI laser alone but the dissociation laser produces some background ions via a one-photon transition to the $\tilde{B}\ ^2A_1'$ state of $CH_3$ followed by absorption of a second photon ionizing $CH_3$.[24] A background is also present on the Br* image as the $\tilde{A} \leftarrow X$ band of $CH_3Br$ is excited not only by the photolysis laser at 215.9 nm but also by the REMPI laser at 238.6 nm. In each case, background contributions were recorded independently and carefully subtracted. Typically these sources of background were around 20% of the signal for methyl detection and 50% of the signal for Br* detection. Table I summarizes the resonant states and the corresponding REMPI wavelengths used for all species detected in this experiment.

The spatial overlap between the laser and molecular beams was optimized using an NO molecular beam and non-resonant multiphoton ionization. To ensure detection of all photofragments, the typical delay introduced between the two lasers did not exceed 5 ns. Both laser polarizations were parallel to the detector plane in order to maintain cylindrical symmetry required for the inverse Abel transformation. Reconstruction of the 3D-velocity map led to fragment kinetic energy and angular distributions. The dissociation energy assumed in this work was reported by Janssen et al. to be $D_0$=2.901+/-0.016 eV.[19]

## III RESULTS AND DISCUSSION

Velocity map images are highly structured, reflecting two different photochemical pathways in $CH_3Br$. The first reflects the photochemistry of the $\tilde{A} \leftarrow X$ band in methyl bromide neutral. The second is due to the photodissociation of $CH_3Br^+$ around 330 nm. Both of these appear to exhibit rich electronically non-adiabatic behavior as discussed in part A for the $CH_3Br$ and in part B for its cation.

### A- Methyl Bromide

## Experimental Images

Figure 1 shows typical images obtained for m/z=15 ($CH_3^+$) and m/z= 79 ($Br^+$). Image 1-(a) shows a one-color experiment using only the photolysis laser at 215.9 nm and is representative of some of the general observations of this work. Here one sees two sharp concentric rings at the outer edge of the image, which arise when the methyl fragment is unintentionally ionized via the R-branch of the $\left(\tilde{B}\ ^2A_1^{'},\ 0_0^0\right)$ transition. The two outer rings correspond to the two Br spin-orbit channels in $CH_3Br$ photolysis at 215.9 nm. For all intents and purposes, this image is background and no attempt was made to analyze its information content. Rather, this background is subtracted from methyl images obtained in two-color experiments. Fig. 1 (b-e) show these results after background subtraction. Here, one also sees two concentric rings at the perimeter of the image, which reflect the velocity and angular distributions of specific vibrational states of $CH_3$ produced in the 215.9 nm photolysis of $CH_3Br$. Here $CH_3$ is detected by 2+1 REMPI via the $3p\ ^2A_2^{''}$ Rydberg state.[25] One also sees a sequence of features closer to the center of the images, which arise from $CH_3^+$ produced in photodissociation of $CH_3Br^+$, which is itself produced by 2-photon non-resonant ionization at 215.9 nm. We return to a discussion of this process later in part B.

The images shown in Fig.1 (b-e) have been recorded with the REMPI laser tuned to peak of the sharp Q-branches of the $\left(3p\ ^2A_2^{''}\ ,\ 2_v^v\right)$ system. Images recorded with or without scanning over the Q-branch profile were found to be identical, indicating that no $CH_3$ rotational selectivity is possible under our conditions. Two color images of $Br(^2P_{3/2})$ and $Br(^2P_{1/2})$ fragments, ionized by 2+1 REMPI, are also shown in Figure 1-(g) and 1-(h) respectively.

As $v_2$ umbrella excitation increases in these experiments, the signal intensity decreases due to a faster predissociation of $CH_3$ in the intermediate $3p\ ^2A_2^{''}$ state used for REMPI detection.[26] Consequently, the origin band image of $CH_3$ (Fig. 1-(b)) is an average over 12000 laser shots compared to 36000 for the others. This predissociation combined with the other uncertainties associated with use of a structureless Q branch for REMPI detection make it impossible to derive the $CH_3$ $v_2$ vibrational population distribution based on intensities of the images. However, if we assume that the energy distribution of bromine fragments reflects only umbrella mode activity, we can reconstruct the $v_2$ vibrational distribution. To test this assumption we probed other vibrational states, for example the symmetric CH stretch mode $\nu_1$. Normally, the excitation of symmetric the CH stretching mode, $\nu_1$ (372 meV), cannot be probed via the Q branch of the $1_1^1$ band (at 333.9 nm) as it overlaps the $P(4)0_0^0$ band.[10,23,27] However, the translational energy released in the simultaneously detected channels is large enough that they can be identified and separated in this imaging experiment. The excess energy for the methyl fragment excited by one quantum of $\nu_1$, following dissociation at 215.9 nm, is 2.01 eV for the Br* channel and 2.47 eV for the Br one. In contrast, the excess energy for the production of $CH_3(v=0)+Br^*$ is 2.385 eV. In this region of the image, each pixel corresponds to about 30 meV of translational energy. Images recorded at the peak of the $1_1^1$ Q-branch (Fig. 1f) show no significant difference compared to those recorded in the $0_0^0$ Q-branch. We also tried to probe the $\nu_1$ vibration through the $1_1^0$ line, without success. Therefore we conclude that the dissociation dynamics do not significantly involve the symmetric stretching mode in $CH_3Br$, in contrast to the photodissociation of $CH_3I$ in which as much as 10% of the internal energy is channeled into the $\nu_1$ mode, albeit only for the ground spin-orbit channel[23,28,29]

This statement seems consistent with the reported absence of $\nu_1$ activity in experiments in the range 212.8 nm to 235 nm detected by slicing at a velocity resolution three times higher than ours.[19] With this high-resolution slicing, Janssen et al.[19] were able to detect fragments excited in the $\nu_4$ rocking mode for $CD_3Br$. It is, however important to remember that this same slicing experiment shows a strong effect of the *D*-isotope substitution enhancing production of vibrationally excited methyl in contrast to the lighter isotope, $CH_3Br$.[19]

The absence of $\nu_1$ activity in $CH_3Br$ dissociation is a somewhat different than $CH_3I$ photodissociation, where the stretching activity observed mainly for the I-channel is associated with impulsive behavior of the photodissociation along the $^1Q_1$ surface at high available energy. This impulsive feature is expected to be stronger in $CH_3Br$ since the $^1Q_1$ surface is steeper than in $CH_3I$.[14] However none of the investigations on $CH_3Br$ or $CH_3Cl$ photodissociation has reported a $\nu_1$ activity. Trajectory studies involving potential energy surfaces that take into account this coordinate might be helpful in explaining this difference observed in the $\nu_1$ activity of the methyl halides.

**Energy Distributions and Correlation of $CH_3$ $\nu_2$ excitation with Br S.O. state**

Fig. 2 shows the kinetic energy probability distributions obtained from analysis of the images of Fig. 1. This experiment yields the kinetic energy of the $CH_3$ fragments as a function of their umbrella excitation as well as the Br and $Br^*$ energy distributions. This energy mapping of the fragments allows a more detailed study of the non-adiabatic coupling. In each translational energy distribution for $CH_3(v)$, the lower kinetic energy peak corresponds to production of $Br^*+CH_3(v,J)$ while the higher energy one results from $Br+CH_3(v,J)$. We fitted each peak with a Lorentzian function. From the derived width parameters (shown in Table I) one can see the $CH_3$ fragments' translational energy distribution is at least ~30 % wider when formed with Br atoms in comparison to those formed with Br*. This broadening can be explained by a higher degree of rotational excitation of the methyl fragment in the more energetic Br-forming channel, similar to the results of Ref. 19.

We calculated the [Br*]/[Br] branching ratio for each vibrationally state specific channel and found that it decreases as the excitation of the $CH_3$ umbrella mode increases. The decreasing [Br*]/[Br] ratio seen here is in agreement with the propensity previously inferred from Br-atom images by Gougousi et al.[5] and Underwood et al.[20]

Under the apparently valid assumption that only $CH_3$ umbrella motion is excited, the sum of individual methyl distributions should reflect closely the bromine fragment distributions. Consequently, we independently fitted the translational energy distributions of the $Br\left({}^2P_{3/2}\right)$ and $Br^*\left({}^2P_{1/2}\right)$ fragments, using the component vibrational state specific translational energy distributions of the $CH_3$ fragments and varying the relative contribution of each $CH_3$ vibrational state.

The vibrational distributions obtained in this way are listed in the last two columns of table I and the fits to the Br and $Br^*$ distributions are displayed on Fig. 3. Here, the vibrational distribution of the $Br^*$ channel peaks at $v_2$=1 and contributions from vibrational state $v_2$=2 and lower are enough to fully reconstruct the distribution. For the Br channel, it is evident that contributions from the first three quanta in the umbrella mode (observed directly in this work) are not enough to reproduce the translational energy distribution; larger $CH_3$ internal energy is clearly required. We were able to fit the Br translational distribution with reasonable assumptions about the vibrational state specific $CH_3$ translational energy distributions for higher values of $v_2$. For contributions from $v_2$ = 4, 5 and 6, we assume the same rotational energy distribution as that implied by the translational energy distribution for $v_2$=3 and used the known vibrational energies of these states.[30] It remained then only to vary the contributions from each vibrational state to arrive at the good fit shown in Fig. 3b.

The vibrational excitation is obviously substantially hotter for the Br channel than for the Br* channel with a maximum near $3v_2$. This confirms the “inverted” distribution previously suggested by Gougousi et al.[5] and Van Veen et al.[13] obtained by less direct means. The combination of direct detection of individual quantum states of $CH_3$ as well as Br and $Br^*$ in this work provides the most reliable and accurate information on the $CH_3$ excitation in $CH_3Br$ photodissociation and how it varies between the two spin orbit states.

**The Br/Br$^*$ branching ratio**

The Br and Br* photofragment quantum yields are related to the [Br*]/[Br] branching ratio by the following equations:

$$\Phi_{Br} = \frac{[Br]}{[Br]+[Br*]} = \frac{1}{1+\frac{[Br*]}{[Br]}} \quad \text{and } \Phi_{Br^*} = 1-\Phi_{Br} \qquad \text{Eq. 1}$$

Since the two-photon transition strength is not known for the Br and Br* REMPI detection scheme we used, it is not possible to evaluate the overall branching ratio [Br*]/[Br] simply by dividing the areas under the Br and Br* fragment translational energy distributions in Fig. 2. However $\Phi_{Br*}$ has been measured by Gougousi et al.[5] down to 218 nm. Extrapolating those results, we find $\Phi_{Br*}$= 0.58±0.05 at 215.9 nm. This leads to a branching ratio of $\frac{[Br*]}{[Br]}$=1.38±0.15 and to $\Phi_{Br}$=0.42±0.05.

The [Br*]/[Br] ratio could depend on the $CH_3$ rotation and vibration. We measured images probing through the $0_0^0$ Q-branch, which show an increase in the [Br*]/[Br] ratio as one tunes from the blue to the red. This effect has been previously observed[26] and studied in $CH_3I$.[31] The bending mode of the parent molecule is not only the promoting mode for the non-adiabatic coupling between $^1Q_1$ and $^3Q_0$, but also correlated with rotation of $CH_3$ fragments. For example, for trajectories that explore bent geometries near the $^1Q_1$-$^3Q_0$ curve crossing, larger rotational excitation of the methyl fragment is expected for the $^1Q_1$ than for the $^3Q_0$ state.[16] Note that from the ratio $\frac{[Br*]}{[Br]}$ in the detection of $CH_3$ produced in the origin band (see Table II), we can compare the yield $\Phi_{Br^*}$ = 0.82±0.04 measured at 215.9 nm in our case on a supersonic molecular beam to $\Phi_{Br^*} = 0.72$ measured at 213 nm by selecting the initial parent molecule (K=J=1) Lipciuc and Janssen.[19] These two ratios are in good agreement.

## Angular distributions and alignment in $CH_3$

The angular distributions, $I(\theta)$, of the photofragments were obtained by integrating the Abel-inverted images over the radial full-width half maximum of each of the peaks shown in Fig. 2. The angular distributions were fitted to the expression:

$$I(\theta) \propto 1 + \frac{\beta}{2}\left(3\cos^2\theta - 1\right) \qquad \text{Eq. 2}$$

where β is the photofragment recoil anisotropy parameter for a given dissociation channel. The distributions and fits are presented in Figure 4 and the values for the anisotropy parameters shown in the first two rows of Table II. Looking at the symmetry of the states involved, one would expect, in the limit of a prompt dissociation, that Br and $CH_3$ produced by excitation to $^1Q_1$ and $^3Q_1$ states to have a perpendicular transition character and $\beta = -1$, while Br* and $CH_3$ produced by excitation to the $^3Q_0$ state to exhibit a parallel transition character and $\beta = +2$.

The β parameters derived from bromine fragment angular distributions are 1.88 for the Br* producing channel and –0.11 for the Br channel, in good agreement with previous results.[5] The $Br^*$ channel's anisotropy agrees approximately with the parallel transition picture of $^3Q_0$ excitation of $CH_3Br$. The Br channel's almost isotropic angular distribution is, however, inconsistent with a perpendicular transition for $CH_3Br$ excitation to $^3Q_1/^1Q_1$ states. This can be explained[5] by the non-adiabatic coupling between the $^3Q_0$ to the $^1Q_1$ state which gives parallel character to the otherwise perpendicular transitions. The observed β parameters can be rewritten as a combination of direct and indirect contributions as follows:

$$\beta(Br) = a_{indirect}\beta(^3Q_0) + a_{direct}\beta(^XQ_1)$$
$$a_{indirect} + a_{direct} = 1$$
$$\beta(Br^*) = b_{direct}\beta(^3Q_0) + b_{indirect}\beta(^XQ_1) \qquad \text{Eq. 3}$$
$$b_{indirect} + b_{direct} = 1$$

Here the coefficient $a_{direct}$ is the quantum yield for direct excitation to the $^XQ_1$ states and dissociation yielding Br($^2P_{3/2}$) atoms. The $a_{indirect}$ coefficient is the probability for excitation to the $^3Q_0$ state followed by non-adiabatic transition and dissociation along the $^XQ_1$ state(s) to form Br($^2P_{3/2}$) atoms. *b*'s are defined in an analogous fashion. The four coefficients of Eq. 3 can be calculated taking the approximation of a pseudo-diatomic system with a dissociation time occurring on a faster timescale than the rotational period of the parent molecule, so that $\beta\left(^3Q_0\right)$ and $\beta\left(^XQ_1\right)$ have the values of a perfect parallel and perpendicular transition respectively, namely $\beta\left(^3Q_0\right)$ = 2 and $\beta\left(^XQ_1\right)$ = −1. The coefficients for Br fragments (Table 2) show that $a_{indirect}$=0.3 of the Br fragments come from $^3Q_0 \rightarrow {}^XQ_1$ non-adiabatic coupling in agreement with results published by Gougousi et al.[5] If the same procedure is applied to the Br* channel, we find that almost all ($b_{direct}$=0.96) Br* is produced by a direct pathway. The excitation probability to different surfaces ($P_{^3Q_0}$ and $P_{^XQ_1}$) and the coupling between these surfaces ($P_{10}$ and $P_{01}$) can be calculated from the quantum yields $\Phi_{Br*}$=0.58±0.05 and $\Phi_{Br}$=0.42±0.05 [5] and *a* and *b* coefficients as follows and the results are summarized in Table II :

$$P_{01} = P_{^3Q_0 \rightarrow {}^XQ_1} = \frac{a_{ind.}\Phi_{Br}}{a_{ind}\Phi_{Br} + b_D\Phi_{Br*}}$$
$$P_{^3Q_0} = a_{ind}\Phi_{Br} + b_D\Phi_{Br*}$$

and

$$P_{10} = P_{^XQ_1 \rightarrow {}^3Q_0} = \frac{b_{ind.}\Phi_{Br*}}{b_{ind.}\Phi_{Br*} + a_D\Phi_{Br}}$$
$$P_{^XQ_1} = b_{ind.}\Phi_{Br*} + a_D\Phi_{Br}$$

Conservation of momentum dictates that the state specific angular distributions of the $CH_3$ fragments, weighted by the $CH_3$ quantum state population distribution, match those of the Br fragments. It is evident from the values in table II that this is not the case. We attribute these differences to alignment of the methyl fragment, which appears to be particularly important for vibrationally excited $CH_3$, as well as the influence of vibrationally enhanced predissociation in the REMPI detection used here.

Rotational alignment has been observed before for methyl following photodissociation of $CH_3I$. Powis and Black[28,32] found that $CD_3$ produced by dissociation of $CD_3I$ at 266nm is strongly aligned, rotating mainly around its $C_3$ axis with K=±N, where N is the total angular momentum excluding spin and K its projection along its three-fold symmetry axis. However this alignment probably originates from rotationally hot parent molecules present in the 300K effusive molecular beam used in that work. Houston and coworkers[26] used a supersonic molecular beam and their experimental results on both $CD_3I$ and $CH_3I$ conclude that $CD_3$ and $CH_3$ fragments are also aligned; however, here K=0 dominates reflecting rotation around an axis perpendicular to the three-fold symmetry axis. These results were confirmed by Janssen et al.[33] who, following the treatment of Kummel and Zare,[34] measured the alignment parameters from O, P, R and S branch methyl spectra in $CD_3I$. In the current case we do not have enough polarization geometries and wavelengths to address this problem quantitatively.

Still we can take an approach which is at least sufficient to demonstrate that alignment effects are present. In an approximate fashion, we fitted our experimental angular distribution to a series expansion of even Legendre polynomials[35] terminated at the sixth order polynomial:

$$I(\theta) = 1 + \beta_2 P_2(\cos\theta) + \beta_4 P_4(\cos\theta) + \beta_6 P_6(\cos\theta) \qquad \text{Eq. 6}$$

In this expansion, alignment will manifest itself with nonzero values for $\beta_4$ and/or $\beta_6$.

The results of this fitting procedure are presented in Table 3 and the fits to the data are presented in Fig. 4. Values of $\beta_6$ are -within our experimental error- zero. However, $\beta_4$ clearly exhibits non-zero values for most vibrational quanta and are generally higher in $CH_3$+Br* than in $CH_3$+Br channel. This result shows the presence of $CH_3$ alignment in both channels and that $CH_3$ alignment is more important in the $CH_3$+Br* channel, although the rotational temperature seems to be lower in that channel. It is likely that the stronger alignment in the $CH_3$+Br* channel are related to its direct dissociation mechanism.

Fitting Br and Br* angular distributions in the same fashion does not indicate significant alignment of Br. Note also that $Br(^2P_{1/2})$ with J=1/2 cannot have alignment.

The presence of alignment as well as the predissociation of the $CH_3$ 3p intermediate Rydberg state[28,32] prevent us from using methyl angular distributions to extract curve crossing probabilities as a function of methyl fragment excitation. However, it is noteworthy that the methyl angular distributions vary only weakly with vibrational excitation (See table III) indicating that the curve crossing probability will not change significantly. This is consistent with the analysis of Underwood et al..[20]

**B- Methyl bromide cation**

**<u>Identification of Ion Dissociation</u>**

In order to determine the origin of the low velocity rings in the $CH_3^+$ images we have also obtained ion images at m/z = 94-96 ($CH_3Br^+$). These images (not shown) appear as tiny dots at the velocity origin, as expected for ionized $CH_3Br$ travelling with the velocity of the molecular beam. When the laser light used for $CH_3$ REMPI detection (hereafter referred to as "the ion-dissociation laser") is introduced at wavelengths near 330 nm, we observe that the $CH_3Br^+$ peak is reduced by as much as ~75% without any significant change in the intensity of $^{79,81}Br^+$. This clearly shows the occurrence of the following channel:[36]

$$CH_3Br + 2\,h\nu_{\lambda=215.9nm} \rightarrow CH_3Br^+ + e^-$$

$$CH_3Br^+ + h\nu_{\lambda=322\text{-}334\text{ nm (“ion-dissociation laser”)}} \rightarrow CH_3^+ + Br/Br^*$$

See Fig. 5 for an energy diagram. The additional smaller-radius rings in the $CH_3^+$ images (Fig.s 1 c-f) correspond to this second source of $CH_3^+$.

The ground electronic state ($^2E_{3/2}$) of $CH_3Br^+$ is formed by the removal of a nonbonding (2e) electron from one of the halogen lone pairs. This state's geometry is similar to the neutral ground state but with a slight elongation of the C-Br bond[37] as well as a Jahn-Teller distortion from $C_{3v}$ geometry.[2] The ionization energy to form $CH_3Br^+$ ( $X\ ^2E_{3/2}$) is 10.54 eV[38,39] and the spin orbit excited $X\ ^2E_{1/2}$ state lies at 10.86 eV.[40] Two photons at 215.9 nm (11.48 eV) are sufficient to reach the ionization continuum but not enough to reach dissociative ionization threshold (12.74 eV) leading to $CH_3^+\left(^1A_1\right)+Br\left(^2P_{3/2}\right)$.[39,41] Although it has never been observed,[42] the appearance potential of $CH_3^+\left(^1A_1\right)+Br\left(^2P_{1/2}\right)$ is expected at 13.2 eV. One photon from the ion-dissociation laser is required to dissociate $CH_3Br^+$. Photoelectron spectroscopy of $CH_3Br$ shows that the first electronically excited state of $CH_3Br^+$ lies 13 eV above $CH_3Br$ with a vertical excitation energy of 13.5 eV.[2,44,45] See Fig. 5. This state is formed by removing an electron from the ($3a_1$) outer-valence molecular orbital associated with the C-Br bond and labeled $A^2A_1$. Excitation to this state yields, with 100% efficiency, the $CH_3^+$ fragment.[44] Some vibrational bands have been observed around 12.8 eV revealing a bound character of the $A^2A_1$ state at energies lower than those relevant to this work.[39]

**The observation of highly vibrationally excited $CH_3^+$**

The inner ring structure observed in Fig.'s 1c-f depends only weakly on the wavelength of the ion-dissociation laser. Fig. 6 shows the translational energy distributions for $CH_3Br^+$ photodissociation derived from each image. In order to compare results at different photolysis wavelengths, the translational energy scale has been shifted relative to the available energy for the experiment carried out with the ion-dissociation laser at $h\nu_{ref}$=3.84 eV (322.8 nm), corresponding to the image shown in Fig. 1e. The energy scale shift was imposed according to the following formulas:

$$E_T(\nu) = E_{kin}^{CH_3}(measured) \times \frac{m_{CH_3Br}}{m_{Br}} + \Delta$$

$$\Delta = h(\nu_{ref} - \nu)$$

where ν is the frequency of the ion-dissociation laser. This approach to the analysis allows us to identify vibrational progressions of the $CH_3^+$ product that appear in different experiments. Using this analysis, we discover a progression of states at center-of-mass translational energies: 33, 190, 376, 528 and 686 meV. Observation of this progression is additional evidence demonstrating the multiphoton scheme of Fig. 5.

To assign these quantized features requires additional consideration as one cannot *a priori* determine which states of the $CH_3Br^+$ are formed by the two-photon ionization at 215.9 nm. Inspection of Fig. 6 shows that nearly all of the ion signal appears below 1.5 eV. With an ion-dissociation laser photon of $h\nu_{ref}$=3.84 eV, one gets an appearance energy for $CH_3^+$+Br at 12.88 which is in good agreement with the appearance energy of 12.74 eV expected for the A state:

$$CH_3Br^+(^2E_{3/2}) + h\nu_{\lambda=322.8nm} \rightarrow (CH_3^+ + Br)\ (E_T=10.54 + 3.84 - 12.74\ eV=1.64\ eV)$$

We are able to reproduce the progression of Fig. 6 using a vibrational harmonic frequency of $CH_3^+$ $\nu_4$=170 meV (in-plane bend) or $\nu_2$=168 $cm^{-1}$ (umbrella mode).[43] Table 4 shows this analysis more completely. More generally, it is clear that there is a propensity for channeling available energy into $CH_3^+$ vibration and not into translation.

Two-photon threshold photoelectron spectra shows that photoionization onto X $^2E_{3/2}$ dominates.[40] If $A^2A_1$ was the dissociating state accessed directly by $CH_3Br^+$ absorption of a 330 nm photon, the angular distribution would be expected to be perpendicular, yet we observe a parallel process. Thus an indirect dissociation process must be important. For example, the 2nd excited electronic state seen in $CH_3Br$ photoelectron spectroscopy around 14.5 eV possesses E symmetry (B state on Fig. 5).[42] This state correlates adiabatically to $CH_3\left(X\,^2A_2''\right)+Br^+(^3P)$. It has been shown in the case of $CH_3Cl$ that a favorable overlap of the orbitals makes the oscillator strength to the B state (π*←π)much larger than to the A state (σ←π).[47] In the same work, a parallel electronic transition has been observed in an energy range similar to that of the A state absorption.[47] In addition, the coupling between the B and A states of the cation has been proposed to rationalize experimental observations in multiphoton ionization of $CH_3I$.[46]

It is interesting to compare the kinetic energy release distributions of $CH_3^+$ observed in Fig. 6 with the ones recorded by one-photon dissociative ionization[42] : The average translational energy observed in a one photon experiment at hν=14.3 eV (close to the present energy), is ~0.4 eV, [42] which is remarkably similar to the observations of this work.
Thus, internal conversion to the $A^2A_1$ state would involve electron transfer from $CH_3$ to Br, possibly explaining the large vibrational excitation seen in the $CH_3^+$ product.

**IV Conclusion**

We have measured angular and kinetic energy distributions of state selected methyl and bromine fragments using velocity map imaging of single and multiphoton excitation of $CH_3Br$ at 215.9 nm. Our single photon data confirm previous results on differences between Br and Br* channels for methyl production with $v_2$ umbrella mode excitation and show that [Br*]/[Br] branching ratio decreases with umbrella mode excitation. The $v_2$-state specific imaging measurements of $CH_3$ photofragments show that $CH_3Br$ photolysis produces vibrationally hot methyl peaking at $v_2$=1 for the Br channel and at $v_2$=3 for the Br* channel. Although absorption to the Br* producing $^3Q_0$ state dominates the $\tilde{A} \leftarrow X$ band at this wavelength, significant amount of Br product is observed, 18% of which comes from $^3Q_0 \rightarrow {}^XQ_1$ avoided crossing. Alignment effects observed in methyl fragments and predissociation of the intermediate states used for REMPI detection prevented the extraction of non-adiabatic probabilities from methyl data as a function of $v_2$. Qualitatively, based on the overall angular distribution behavior with $v_2$ excitation there seems to be no significant dependence on the methyl umbrella mode excitation on the curve crossing probability.

In addition to single photon data, a 2+1 excitation scheme of $CH_3Br$ is detected in methyl images, where two photons of 215.9 nm light produce $CH_3Br^+$ which subsequently absorbs a photon of 320-330 nm and is excited via a parallel transition to E-symmetry excited ionic state. Coupling of those states to the A excited state of $CH_3Br^+$ leads to production of vibrationally hot $CH_3^+$ (possible in the $\nu_4$ degenerate bending mode excited by the nonadiabatic state mixing) and Br*.

**Acknowledgments:**

The authors would like to thank B. Lepetit, S. Pratt and P. Rakitzis for fruitfull discussions. V.B. acknowledges support from the CNRS institution during her stay in UCSB. AMW and PS acknowledge the support of the AFOSR under contract number FA9550-04-1-0057.

## TABLES

**Table I State specific imaging results.** Intermediate states through which resonant ionization of photofragments has been recorded, Br*/Br yield as shown in Fig 2, width of Lorentzian for the translational energy and vibrational distributions deduced from fits shown in Fig 3.

| | | Detection | Product Internal Energy (in $cm^{-1}$) | [Br*]/[Br] | Lorentzian width (in meV) | | Vibrational population /% | |
|---|---|---|---|---|---|---|---|---|
| | | | | | Br | Br* | Br | Br* |
| Br | | (2+1) REMPI via $^4D^0_{5/2}$ at 215.9 nm | 0 | 1.38±0.15 From [5] | ~466 | - | - | - |
| Br* | | (2+1) REMPI via $^2P^0_{1/2}$ at 238.6 nm | 3685 | | - | 229±17 | - | - |
| $\nu_2$ of $CH_3$ | 0 | (1+1 ) REMPI $0^0_0$ of $\tilde{B}$ $^2A_1'$ At 215.9 nm[24] | 0 | 1.03±0.08 | 257±13 | 163±8 | - | - |
| | 0 | (2+1) REMPI $0^0_0$ of $3p$ $^2A_2''$ At 333.4 nm[22] | 0 | 4.52±0.9 | 157±8 | 122±7 | 0 | 23±3 |
| | 0 | (2+1) REMPI $0^0_0$-P(4) of $3p$ $^2A_2''$ At 333.9nm [10] | 0 | 8.37±1.88 | 102±2 | 95±13 | - | - |
| | 1 | (2+1) REMPI $2^1_1$ of $3p$ $^2A_2''$ At 329.5 nm[23] | 606 | 1.72±0.14 | 172±20 | 108±7 | 16±2 | 46±4 |
| | 2 | (2+1) REMPI $2^2_2$ of $3p$ $^2A_2''$ At 326.1 nm[23] | 1288 | 0.317±0.03 | 169±10 | 133±11 | 20±5 | 31±3 |
| | 3 | (2+1) REMPI $2^3_3$ of $3p$ $^2A_2''$ At 322.8 nm | 2019 | 0.142±0.02 | 175±9 | 125±14 | 21±5 | 0 |
| | 4 | | 2791 | No data | | | 19±4 | 0 |
| | 5 | | 3602 | No data | | | 16±3 | 0 |
| | 6 | | 4445 | No data | | | 8±2 | 0 |

**Table II:** Measured anisotropy parameter, β, shown in Fig 4 and coefficients for direct and non-adiabatic dissociation for both channels. See text.

| Fragment | Br | $CH_3$ for different $\nu_2$ | | | | | |
|---|---|---|---|---|---|---|---|
| | ∑v | 0<br>(1+1)<br>REMPI | 0-P(4)<br>(2+1)<br>REMPI | 0<br>(2+1)<br>REMPI | 1<br>(2+1)<br>REMPI | 2<br>(2+1)<br>REMPI | 3<br>(2+1)<br>REMPI |
| β for<br>Br channel | -0.11±0.01<br>(-0.11±0.02<br>From [5]) | -0.43<br>±0.02 | -0.22<br>±0.02 | -0.096<br>±0.012 | -0.52<br>±0.02 | -0.59<br>±0.03 | -0.62<br>±0.03 |
| β for<br>Br* channel | 1.88±0.06<br>(1.86±0.15<br>From [5]) | 1.35<br>±0.03 | 1.35<br>±0.04 | 1.82<br>±0.04 | 1.54<br>±0.03 | 1.74<br>±0.09 | 1.62<br>±0.12 |

| Br | | | | Br* | | | |
|---|---|---|---|---|---|---|---|
| $a_{indirect}$<br>$^3Q_0 \rightarrow {}^XQ_1$ | $a_{direct}$<br>$^XQ_1$ | $P\left({}^XQ_1\right)$ | $P_{01}$<br>$^3Q_0 \rightarrow {}^XQ_1$ | $b_{direct}$<br>$^3Q_0$ | $b_{indirect}$<br>$^XQ_1 \rightarrow {}^3Q_0$ | $P\left({}^3Q_0\right)$ | $P_{10}$<br>$^XQ_1 \rightarrow {}^3Q_0$ |
| 0.296<br>±0.004 | 0.703<br>±0.004 | 0.32<br>±0.05 | 0.18<br>±0.04 | 0.96<br>±0.02 | 0.04<br>±0.02 | 0.68<br>±0.07 | 0.07<br>±0.05 |

**Table III :** Angular distribution fits to an expansion of Legendre polynomials: Due to more rapid predissociation for higher vibrational states, higher vibrational levels have less intensity and greater experimental uncertainty.

| *Mass* | *State* | $\beta_2$ | $\beta_4$ | $\beta_6$ |
|---|---|---|---|---|
| $CH_3$ Br* | $0_0^0(Q)$ | 1.79±0.01 | 0.16±0.01 | -0.02±0.01 |
| | $0_0^0 P(4)+1_1^1$ | 1.32±0.03 | 0.19±0.03 | 0.09±0.03 |
| | $2_1^1$ | 1.26±0.05 | 0.29±0.05 | 0.03±0.07 |
| | $2_2^2$ | 1.28±0.08 | 0.20±0.08 | 0.09±0.1 |
| | $2_3^3$ | 1.57±0.12 | 0.36±0.10 | -0.20 ±0.13 |
| Br | $^2P_{3/2}$ | -0.11±0.01 | 0.02±0.02 | 0.005±0.02 |
| $CH_3$ Br | $0_0^0 P(4)+1_1^1$ | 0.42±0.08 | 0.18±0.1 | 0.01±0.1 |
| | $0_0^0$(Q) | -0.03±0.01 | -0.08±0.01 | -0.03±0.01 |
| | $2_1^1$ | -0.55±0.05 | 0.08±0.04 | -0.12±0.04 |
| | $2_2^2$ | -0.54±0.01 | -0.09±0.01 | 0.02±0.03 |
| | $2_3^3$ | -0.62±0.02 | -0.22±0.02 | 0.03±0.03 |

**Table IV :** Vibrational progression in $CH_3^+$ fragment

| **Experimentally Derived**[a)] | **v**[b)] **from** $\mathbf{E_{3/2}}$ | **v**[b)] **from** $\mathbf{E_{1/2}}$ | **Fitted Results**[c)] | **(Exp.-Fit)** |
|---|---|---|---|---|
| 686 meV | 3 | 5 | 693 meV | 7 meV |
| 528 | 4 | 6 | 527 | 1 |
| 376 | 5 | 7 | 362 | 6 |
| 190 | 6 | 8 | 197 | 7 |
| 33 | 7 | 9 | 34 | 1 |

a) From observations presented in Fig. 6.

b) Vibrational quanta in $\nu_2$ or $\nu_4$. See text.

c) based on the formula 1282 – [169.9 (v+1/2) +0.468 $(v+1/2)^2$] in meV. The appearance threshold of dissociation is at 1.197+/-0.010 eV in good agreement with $CH_3Br^+(^2E_{3/2}) \rightarrow CH_3^+ + Br^*(^2P_{1/2})$

OR based on the formula 1615 – [169.9 (v+1/2) +0.350 $(v+1/2)^2$] in meV. The appearance threshold of dissociation is at 1.531+/-0.013 eV in good agreement with $CH_3Br^+(^2E_{1/2}) \rightarrow CH_3^+ + Br^*(^2P_{1/2})$

**Figure Captions**

**Fig 1:** Images of methyl radicals (a-f) produced from the 215.9 nm photolysis laser with background (a) subtracted. The images were respectively acquired over 12000 (b) and 36000 (a,c-f) laser shots. Images of bromine fragments produced in (g) Br and (h) Br* states. The images are presented in 0-255 grey scale with the darkest shade corresponding to the maximum of the signal.

**Fig 2:** Kinetic energy release distributions. The top panel is obtained from bromine fragment images. The others are obtained from specific vibrational states of the methyl fragment detected via the $\left(3p\ {}^2\mathrm{A}_2^{"}\ ,\ 2_{\mathrm{v}}^{\mathrm{v}}\right)$ state. The vertical dotted lines indicate the expected maximum kinetic energy release in the methyl fragments if no methyl rotation were excited. $Br\left({}^2P_{3/2}\right)$ results have been derived from images recorded at two repeller voltages : 3kV (diamonds) and 5 kV (circles).

**Fig 3:** Reconstruction of Br translational energy distributions from vibrationally state specific translational energy distributions of $CH_3$. (a) $Br^*\left({}^2P_{1/2}\right)$ channel and (b) $Br\left({}^2P_{3/2}\right)$ channel. See text. Diamonds correspond to images recorded at 3 kV for the repeller plate and crossed circle at 5 kV.

**Fig 4**: Angular distributions of bromine atom and methyl fragment calculated from the images shown in Fig. 1. In each panel the circles correspond to the distribution for the Br channel and the squares correspond to the Br* channel. The solid lines corresponds to the fit using Eq. 2 with the anisotropy parameters listed in the Table II.

**Fig 5**: Energy diagram of the photodissociation of methyl bromide cation in $C_{3v}$ geometry with adiabatic correlation to products shown by dotted lines.

**Fig 6: Vibrational Progression of $CH_3^+$ obtained from central rings in the images of Fig. 1.** An energy shift, Δ, relative to the photodissociation of $CH_3Br^+$ taking place at 322.8 nm ($2_3^3$ REMPI wavelength) is introduced to account for the change in photolysis wavelength in each image. The maximum available energy for translation is indicated by vertical lines, for photodissociation of $CH_3Br^+$ to $CH_3^+$+Br or $CH_3^+$+Br* with a cation initial state $X\,^2E_{3/2}$ (solid lines) or $X\,^2E_{1/2}$ (dashed lines).

V. Blanchet et al., Fig. 1

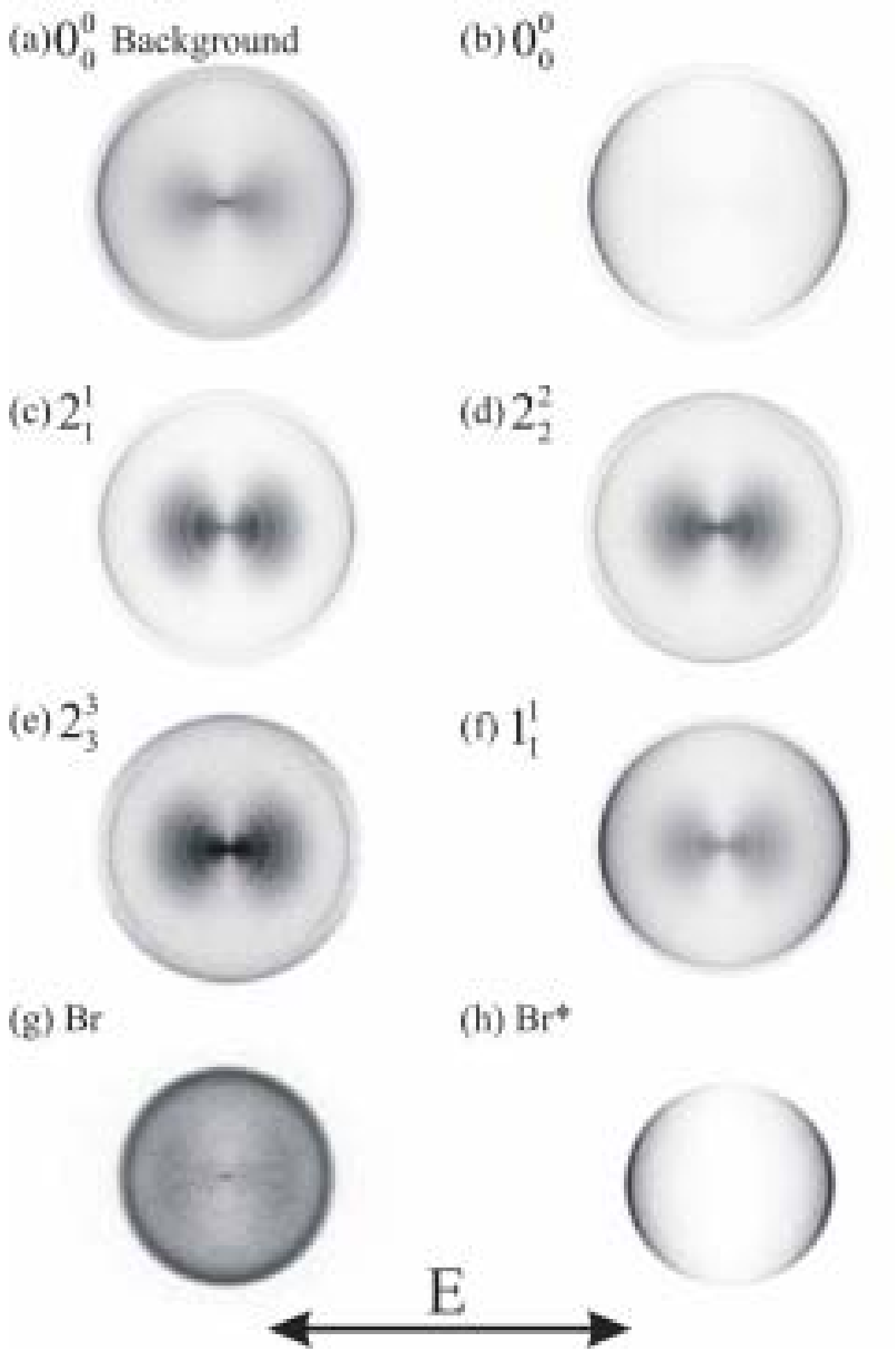

V. Blanchet et al., Fig. 2

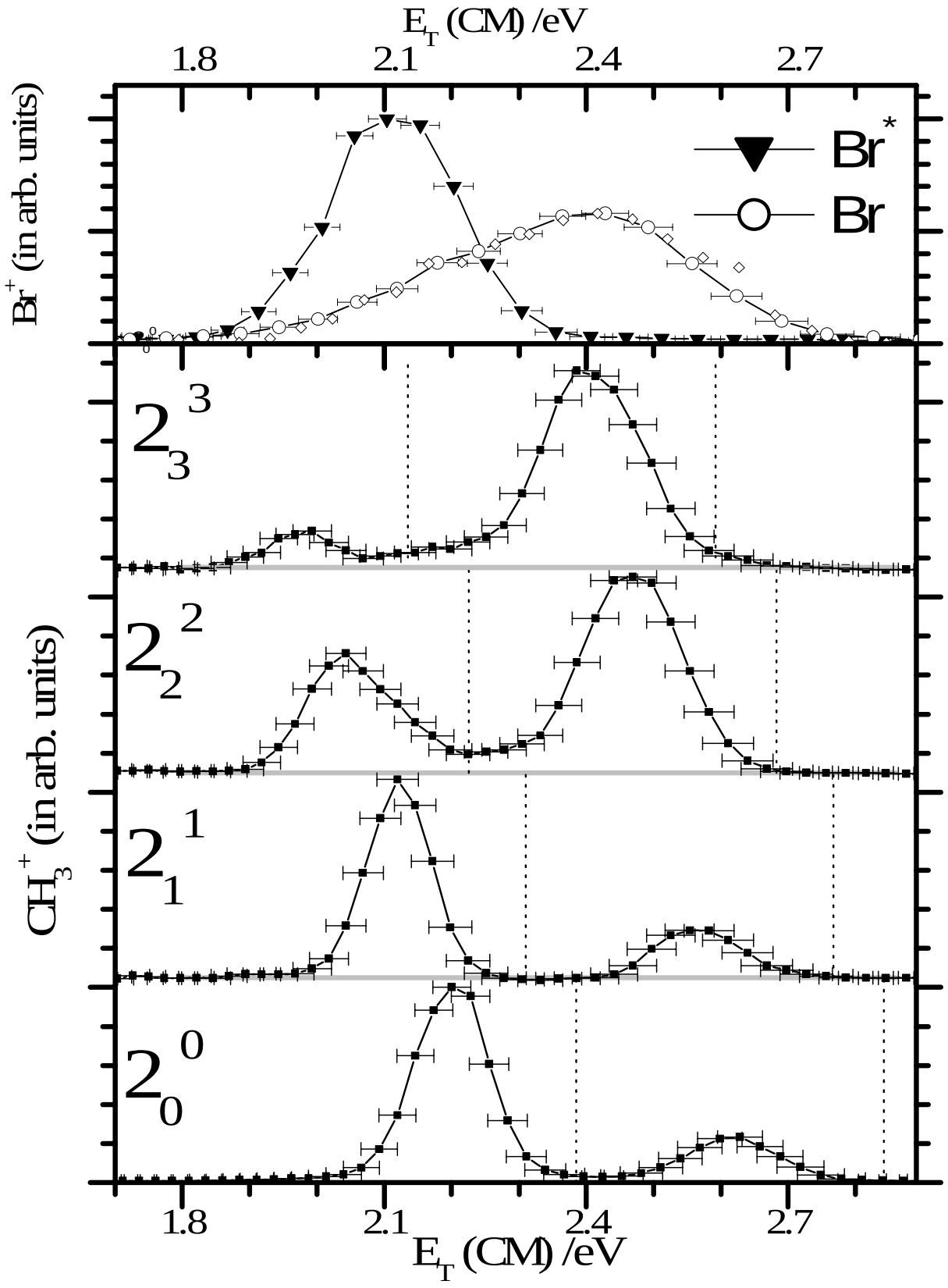

V. Blanchet et al., Fig. 3

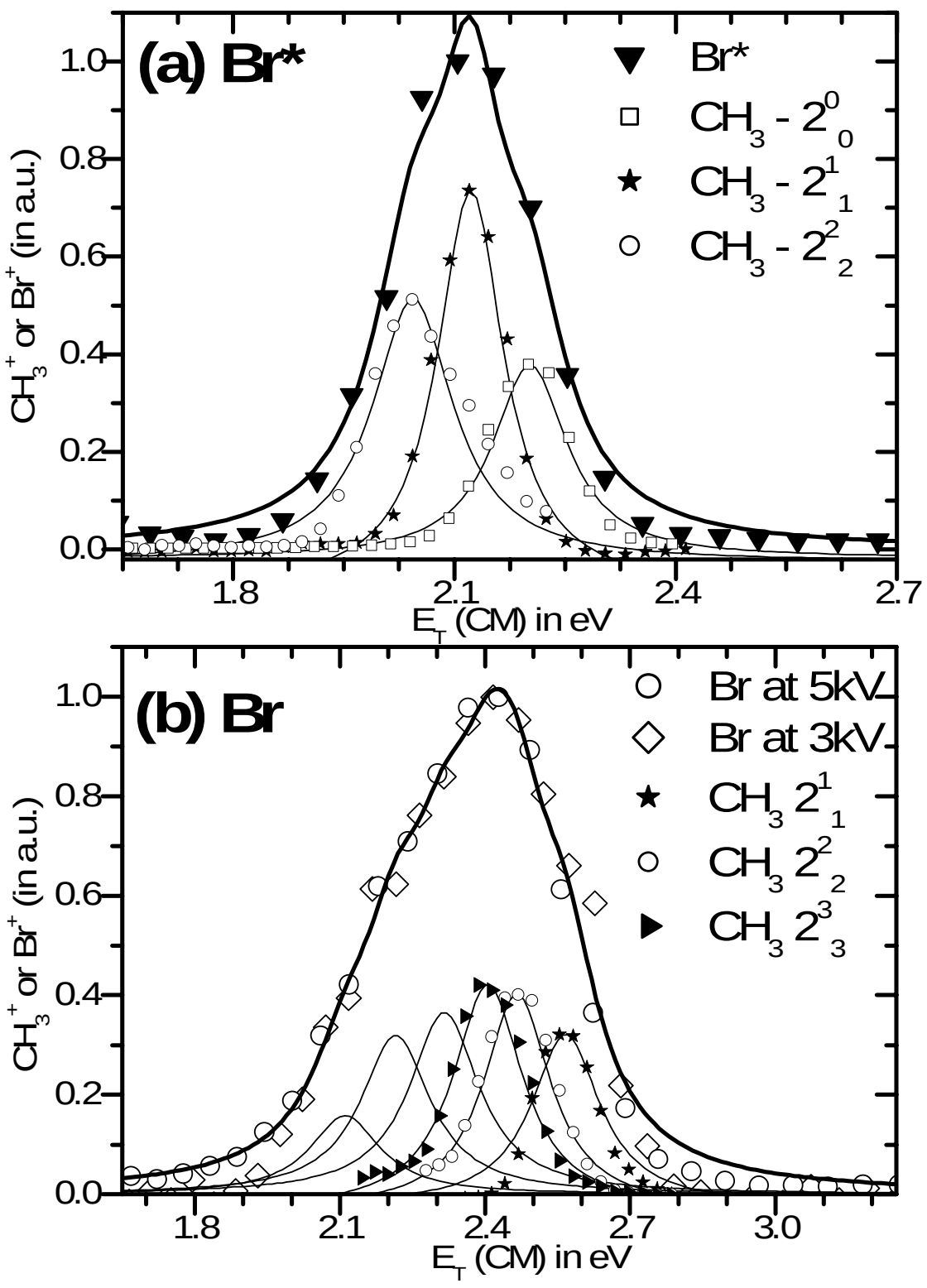

V. Blanchet et al., Fig. 4

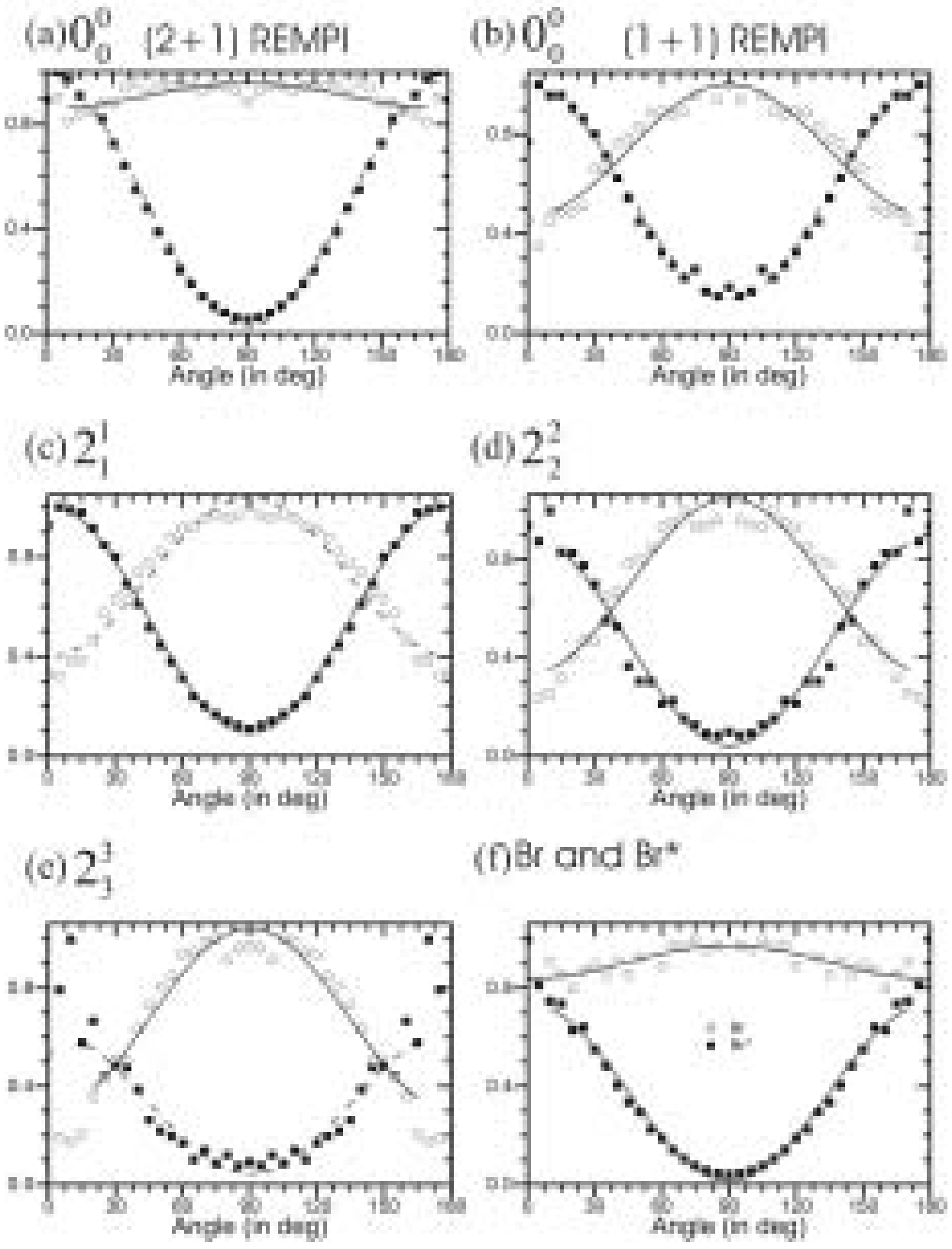

V. Blanchet et al., Fig. 5

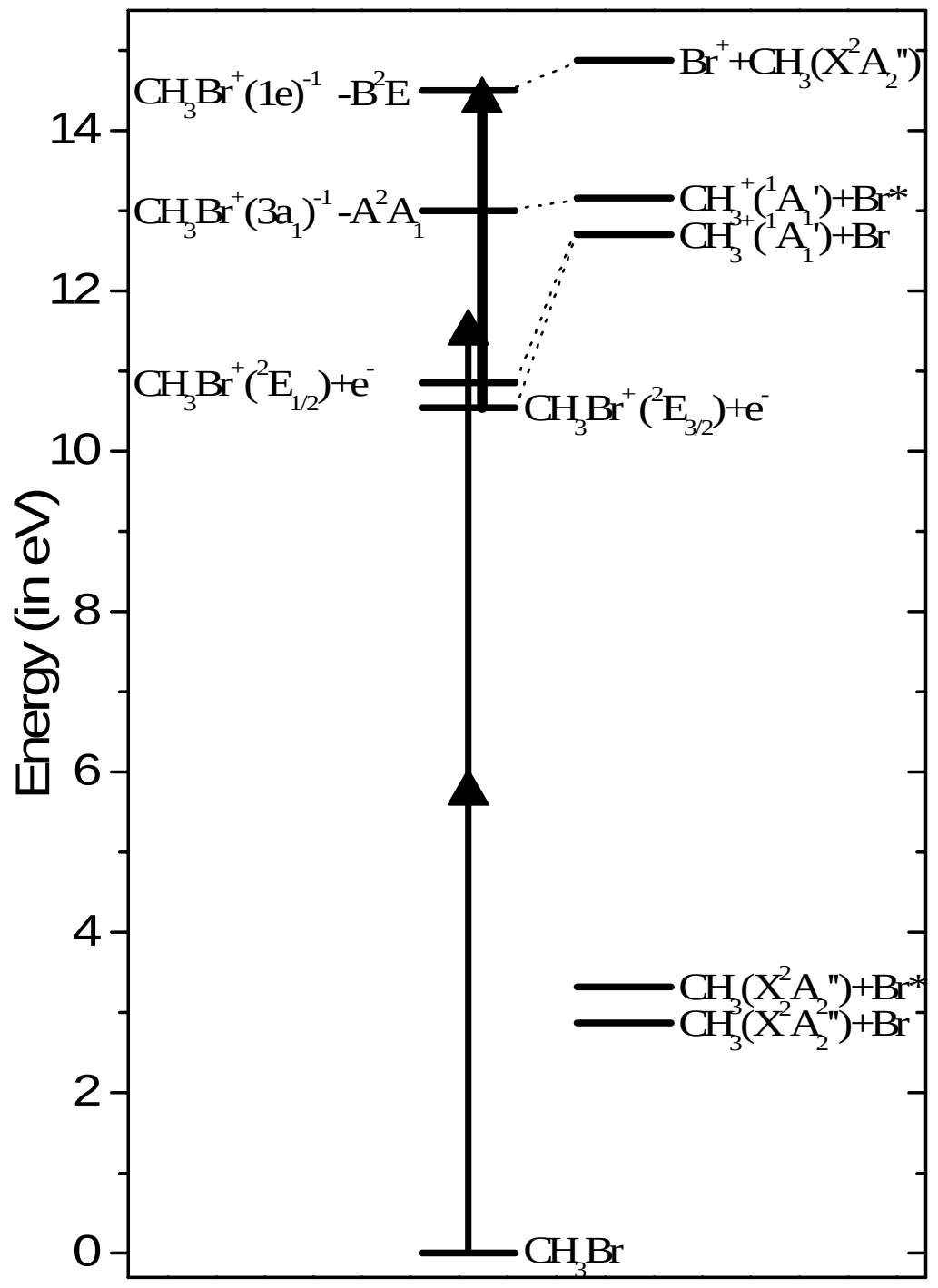

V. Blanchet et al., Fig. 6

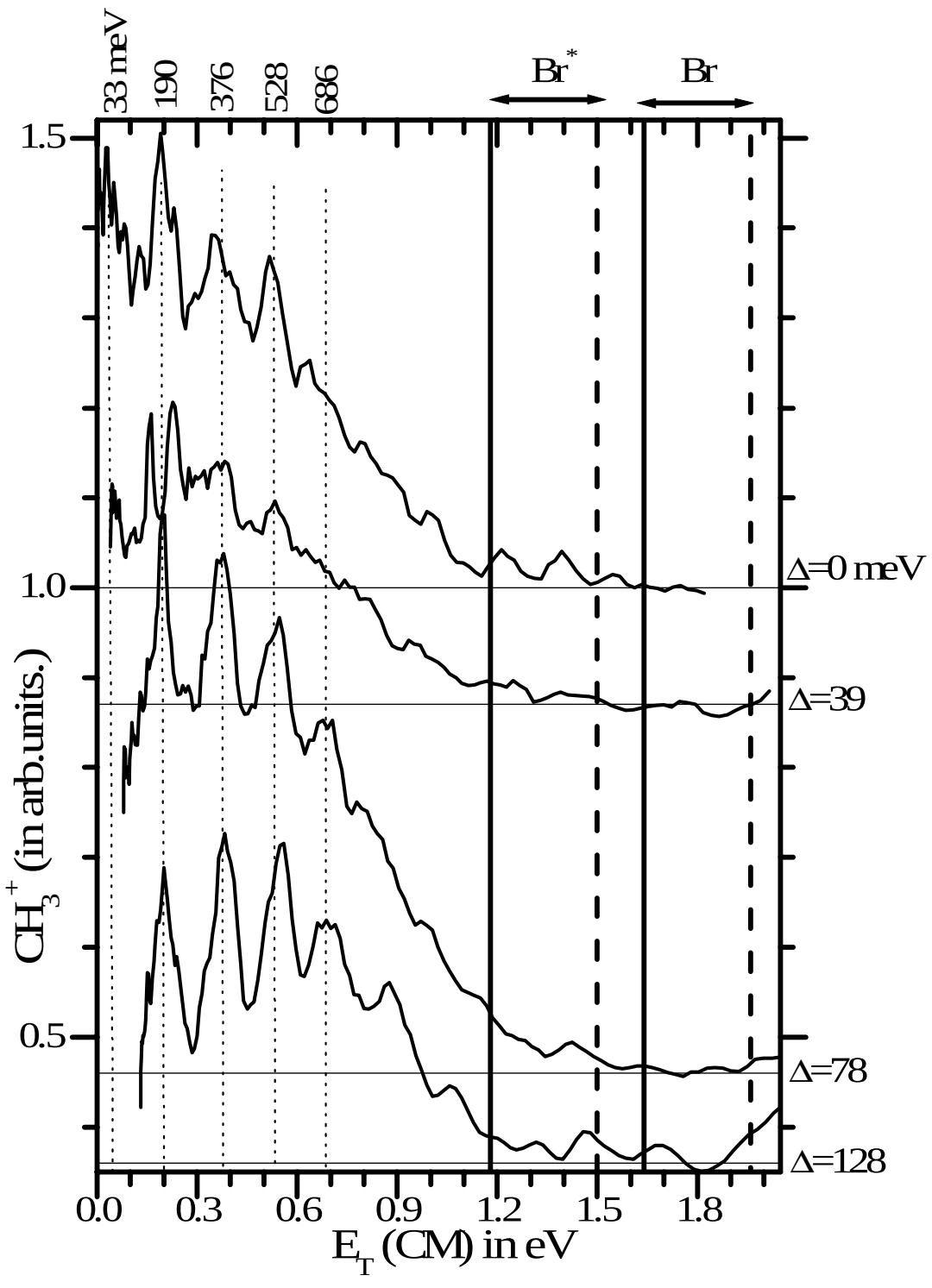